\documentclass[twocolumn,aps,showpacs,floatfix,pra,10pt]{revtex4-1}
\usepackage{amssymb,amsmath}
\usepackage{hyperref,graphicx}
\usepackage{dcolumn}
\usepackage{color}
\usepackage{ulem}
\usepackage{bbold}

\begin{document}

\title{Statistical approach to Casimir-Polder potentials in heterogeneous media}

\author{Nicolas Cherroret, Romain Gu\'erout, Astrid Lambrecht and Serge Reynaud}                          
\affiliation{
\mbox{Laboratoire Kastler Brossel, UPMC-Sorbonne Universit\'es, CNRS, ENS-PSL Research University, Coll\`{e}ge de France,}\\
\mbox{4 Place Jussieu, 75005 Paris, France}
}  

\begin{abstract}
We explore the statistical properties of the Casimir-Polder potential between a dielectric sphere and a three-dimensional heterogeneous medium, by means of extensive numerical simulations based on the scattering theory of Casimir forces. The simulations allow us to confirm recent predictions for the mean and standard deviation of the Casimir potential, and give us access to its full distribution function in the limit of a dilute distribution of heterogeneities. These predictions are compared with a simple statistical model based on a pairwise summation of the individual contributions of the constituting elements of the medium. 
\end{abstract}

\pacs{34.35.+a, 42.50.Ct, 42.25.Fx}

\maketitle

\section{Introduction}

Materials placed in a close vicinity to each other modify the modes of the electromagnetic field. This results in a change of the vacuum energy, which eventually manifests itself as a net force known as the Casimir force \cite{Casimir48, CasimirP48}. The Casimir force has been the subject of a number of experimental investigations at object separations ranging from tens of nanometers to a few micrometers. Starting with the experiments by Lamoreaux
\cite{LamoreauxPRL1997} and Mohideen \cite{MohideenPRL1998}, the Casimir effect
has experienced an enormous increase in experimental activities in recent years 
\cite{EderthPRA2000,ChanPRL2001,ChenPRL2002,DeccaPRL2003,Decca07,vanZwolAPL2008,ChanPRL2008,JourdanEPL2009,Munday09,MasudaPRL2009,deManPRL2009,ChanPRL2010,TorricelliEPL2011,
SushkovNatPh2011,TangPRL2012,DeccaNatComm2013}.

Theoretical approaches to the Casimir force are usually built on an effective medium description of the interacting materials. Within such an approximation, the local details of the materials' microstructure are neglected and the objects are described by macroscopic, spatially-independent dielectric constants. While the effective medium description is in general quite satisfactory for describing dense materials that indeed look homogenous at the typical scales of the Casimir force, this is not necessarily the case for strongly heterogeneous (``disordered'') media that are made of many constituting elements (``scatterers'') well separated from one another. Examples of such heterogeneous systems include nanoporous materials \cite{Pasquini06}, clouds of cold atoms \cite{Schneeweiss12} and, in a slightly different context, corrugated surfaces \cite{Genet03, Neto05}.

From a theoretical viewpoint, interaction phenomena involving strongly heterogeneous materials have been little studied. Seminal works on that subject considered the thermal Casimir interaction between slabs made of piled layers separated from random distances (one-dimensional disorder) \cite{Dean09, Dean10}. The question of disorder was also addressed recently \cite{Allocca15} in the context of the Casimir-Polder (CP) interaction \cite{Intravaia11} between a sphere and a plate \cite{Canaguier11}. In a recent work finally, the CP interaction between a dielectric sphere (or an atom) and a three-dimensional disordered dielectric material was also investigated \cite{Cherroret15}. This is the scenario we consider in the present paper. 

When a probe sphere or an atom interacts with a spatially heterogeneous material such as a semi-infinite disordered medium, the CP potential naturally fluctuates in space. In other words, the Casimir interaction depends on the specific statistical realization of the disorder. A shared conclusion of Refs. \cite{Dean09, Dean10, Allocca15, Cherroret15} is that when the two objects are far enough from each other, precisely when the distance between them is large compared to the typical separation between two heterogeneities, the value of the Casimir potential from a realization to another  is well captured by its configuration average, which coincides with the prediction of the effective medium prescription. In strong contrast, at smaller distances fluctuations of the potential become larger than its mean, which is consequently no longer representative. In practice, this conclusion is crucial for measurements of quantum reflection \cite{Pasquini06, Friedrich02, Voronin05, Dufour13a, Dufour13b}, and more generally for any measurement of the Casimir force involving heterogeneous materials.

In our previous work \cite{Cherroret15}, we developed an exact mathematical treatment of the fluctuations of the CP interaction between a dielectric sphere and a dilute disordered dielectric medium, and applied it to the calculation of the 
mean value of the CP potential and of its standard deviation. In this paper, we consider the same geometry (recalled in Sec. \ref{framework}), for which we perform extensive numerical simulations of the CP potential. The results of these simulations confirm the predictions of \cite{Cherroret15} (Sec. \ref{numerics_sec}), and additionally allow us to compute the full probability distribution of the CP potential which, for a given distribution of the scatterers, does not depend on the microscopic properties of the latter. In a second time (Sec. \ref{theory_sec}), we present a simple statistical model based on a pairwise summation of the individual contributions of the scatterers, and confront it with the simulations. Concluding remarks are collected in Sec. \ref{conclusion_sec}.

\section{Mean and standard deviation of the Casimir-Polder potential}
\label{framework}

\begin{figure}[h]
\includegraphics[width=0.75\linewidth]{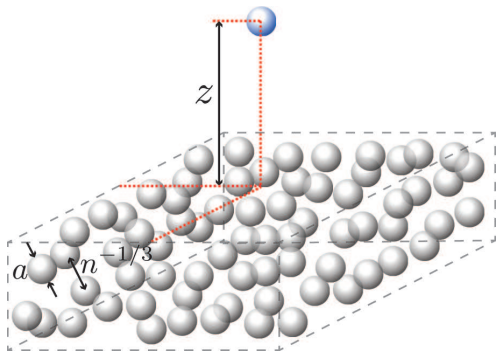}
\caption{
\label{scheme}
(color online) We consider the Casimir-Polder interaction between a dielectric sphere (placed in vacuum) and a semi-infinite disordered medium. The disordered medium consists of a collection of dielectric spheres (size $a$, density $n$) whose positions are uniformly distributed in space.
}
\end{figure}

We address the CP interaction between a probe dielectric sphere (placed in vacuum) of static polarizability $\alpha_0$ (here and in the following, polarizabilities are expressed in SI units divided by $\epsilon_0$) and a semi-infinite, three-dimensional disordered medium consisting of a collection of many scatterers, as illustrated in Fig. \ref{scheme}. We denote by $z$ the distance between the sphere and the surface of the disordered medium. For definiteness, in this paper we restrict our discussion to the retarded regime of the Casimir interaction where $z$ much exceeds the resonance wavelength $\lambda_0$ of the probe sphere (the treatment of the opposite limit $z\ll\lambda_0$ is analogous). Scatterers are also modeled by dielectric spheres of size $a$ and of static polarizability $\alpha_s$. Throughout the paper, we assume that they are uniformly distributed in space with density $n$, and we consider the case of a dilute disordered medium, such that the average distance between the scattering spheres is larger than their size, $n a^3\ll 1$. This is the typical regime where the statistical fluctuations of the CP potential are the largest \cite{Cherroret15}. In the opposite limit $n a^3\sim 1$ of scatterers very close to each other, the approach developed below does not apply but we expect the statistics of the Casimir-Polder potential to be approximately captured by a model where the atom interacts with a rough surface \cite{Moreno10}.

In \cite{Cherroret15},  the question of fluctuations in the limit $na^3\ll 1$ was tackled with the help of a statistical description of the disordered material, in which the CP potential $U(z)$ becomes a random variable. Its mean, $\overline{U}(z)$, and its variance, $\overline{\delta U^2}(z)$, were calculated from an exact treatment of radiation-matter interaction, based on the scattering approach to Casimir forces \cite{Lambrecht06, Emig07} combined with a diagrammatic description of radiation scattering off the disordered medium \cite{Lagendijk96, Rossum99}. In the limit $z\gg\lambda_0$, the following expression for the mean was found \cite{Cherroret15}:
\begin{equation}
\label{Ubar_theo}
\overline{U}(z)=\frac{23}{60}n\alpha_s U^*(z),
\end{equation}
where $U^*(z)=-3\alpha_0 \hbar c/(32\pi^2 z^4)$ is the Casimir potential between the probe sphere and a perfect mirror. As announced, the result (\ref{Ubar_theo}) coincides with the prediction of an effective medium description where the probe sphere interacts with an homogeneous surface of relative permittivity $\tilde\epsilon=1+n\alpha_s$. The amplitude of fluctuations, quantified by the ratio $\gamma$ of the standard deviation of the CP potential and its mean, was found to be (for $z\gg\lambda_0$):
\begin{equation}
\label{gamma_theo}
\gamma=\frac{\sqrt{\overline{\delta U^2}(z)}}{|\overline{U}(z)|}\simeq\frac{a_1}{\sqrt{nz^3}},
\end{equation}
with $a_1\simeq 0.7$. Equation (\ref{gamma_theo}) indicates that $\overline{U}(z)$, i.e. the prediction of the effective medium theory, is well representative of $U(z)$ only when $z\gg n^{-1/3}$. At smaller scales, $\gamma$ becomes larger than unity and $\overline{U}(z)$ no longer provides a trustful estimation of the interaction.

\section{Numerical simulations}
\label{numerics_sec}

\subsection{Methodology}
\label{methodology}

We now propose to investigate the statistical properties of the CP potential from exact numerical simulations in the geometry of Fig. \ref{scheme}. For this purpose, we proceed as follows. We generate an ensemble of $N$ dielectric spheres of radius $a$ and frequency-dependent permittivity $\epsilon(\omega)$, uniformly distributed in a cube of side $L$. This system constitutes a disordered medium of average density $n=N L^{-3}$. An additional probe sphere is placed above this cube, at a distance $z$ to the center of one face, as in Fig. \ref{scheme}.
Denoting by $U^{(N+1)}(z)$ the total, internal Casimir energy between the $N+1$ spheres~\cite{Rahi2009}, the CP interaction $U(z)$ is by definition
\begin{eqnarray}
\label{CP_U_def}
U(z)&=&
U^{(N+1)}(z)-U^{(N+1)}(z \to \infty)\nonumber\\
&=&U^{(N+1)}(z)-U^{(N)}.
\end{eqnarray}
The strategy thus consists in calculating the interaction energy as the difference between the internal energies of $N+1$ and $N$ spheres. Within the scattering formalism~\cite{Lambrecht06,Emig07}, the total Casimir energy between $N$ spheres is given by ~\cite{Rahi2009}
\begin{equation}
\label{multiScatter}
U^{(N)}=\frac{\hbar}{2 \pi}\int_{0}^{\infty}\text{d}\omega
\log \det\left( \mathbb{M}\mathbb{M}_{\infty}^{-1} \right).
\end{equation}
$\mathbb{M}$ is a block-square matrix of dimension $N$ with the following structure:
\begin{equation}
\label{matrixM}
\mathbb{M}=
\begin{pmatrix}
\mathbf{R}_{1}^{-1} &  \mathbf{T}_{1\to 2}  & \ldots & \mathbf{T}_{1\to N}\\
\mathbf{T}_{2\to 1}  &  \mathbf{R}_{2}^{-1} & \ldots & \mathbf{T}_{2 \to N}\\
\vdots & \vdots & \ddots & \vdots\\
\mathbf{T}_{N \to 1}  &   \mathbf{T}_{N \to 2}       &\ldots &\mathbf{R}_{N}^{-1}
\end{pmatrix}.
\end{equation}
The diagonal blocks of $\mathbb{M}$ are the inverse of the spheres' reflection operators $\mathbf{R}_i$. The $(i,j)$ off-diagonal block of $\mathbb{M}$ contains the translation operator $\mathbf{T}_{i \to j}$, which relates an outgoing spherical wave centered on $\mathbf{r}_{i}$  to an incoming spherical wave centered on $\mathbf{r}_{j}$ \cite{Rahi2009}. Finally, $\mathbb{M}_{\infty}^{-1}$ is the block-diagonal matrix $\text{diag}(\mathbf{R}_{1},\ldots \mathbf{R}_{N})$. For the simulations, we express the scattering and translation operators $\mathbf{R}_{i}$ and $\mathbf{T}_{i \to j}$ in a basis of spherical vector waves $| \ell m P \rangle$, with $\ell > 1$, $-\ell \leq m \leq \ell$ and $P=\{\text{E},\text{M}\}$. In this basis, the matrix elements of $\mathbf{R}_{i}$ are given by the standard Mie scattering amplitudes \cite{Canaguier10}. We compute these amplitudes without any approximation, taking into account the full multipole expansion. Finally, we evaluate the matrix elements of $\mathbf{T}_{i \to j}$
 using the formalism of Ref.~\cite{Wittmann1988}.

\subsection{Mean and standard deviation}
\label{av_var_hamaker}

\begin{figure}
\includegraphics[width=0.88\linewidth]{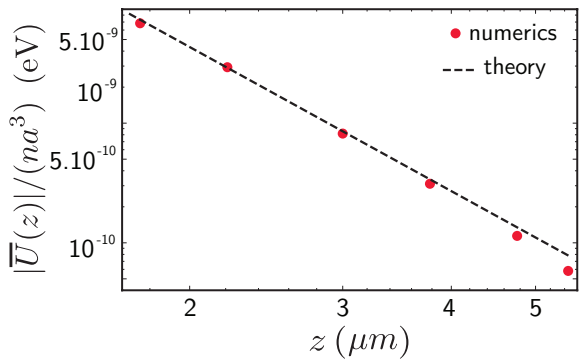}
\caption{
\label{averagez_CP}
(color online) Absolute value of the mean CP potential between a dielectric sphere and a semi-infinite disordered medium, as a function $z$. Dots are the results of exact numerical simulations and the dashed line is the theoretical prediction (\ref{Ubar_theo}).
}
\end{figure}

\begin{figure}
\includegraphics[width=0.8\linewidth]{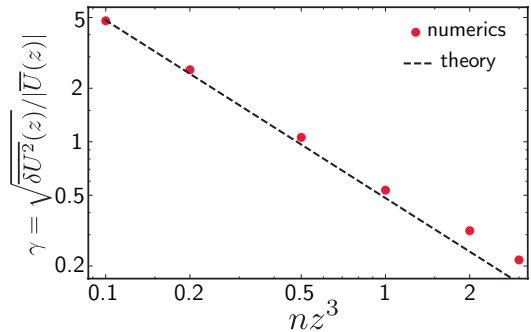}
\caption{
\label{varz_CP}
(color online) Relative fluctuations of the CP potential as a function of $nz^3$. Dots are the results of exact numerical simulations and the dashed line is the theoretical prediction (\ref{gamma_theo}).
}
\end{figure}

Making use of the approach described in Sec. \ref{methodology}, we compute the CP potential $U(z)$ between a dielectric sphere of radius $a=10\,$nm and a disordered medium consisting of $N=32$ other, identical spheres with the same radius $a$, uniformly distributed in a cube of side $L=12\,\mu$m. For these parameters, the disordered medium is dilute, $na^3\ll1$, and we are effectively describing the geometry of a semi-infinite bulk system as long as $z\ll L$. For definiteness, we give to all the spheres the permittivity of silicon: $\epsilon(\omega)=1-[\epsilon(0)-1]\omega_0^2/(\omega^2-\omega_0^2+i\gamma\omega)$, where 
%$\omega_p=2\pi c/\lambda_p$ with $\lambda_p=90.4$nm, 
$\omega_0=2\pi c/\lambda_0$ with $\lambda_0=295$nm, $\epsilon(0)=11.6$ and $\gamma=0.03\omega_0$ \cite{Palik97}. Figure \ref{averagez_CP} displays the absolute value of the disorder-averaged CP potential, $\overline{U}(z)$, computed with these parameters, for several values of $z$ (red dots). 
%Note that for these parameters, $z$ is large as compare to $\lambda_0$ and we are typically probing the retarded limit of the Casimir interaction. 
Each dot is obtained by generating thousands (typically between $3000$ and $8000$ depending on $z$) of disorder realizations, computing $U(z)$ for each of them and finally averaging the results.
In Fig. \ref{averagez_CP}, we also show the theoretical prediction (\ref{Ubar_theo}) (dashed line), which is in very good agreement with the numerics. The small disagreement visible at large $z$ stems from deviations to the geometry of the semi-infinite medium: when $z$ becomes of the order of $L/2$, the probe sphere starts to be sensitive to the boundaries of the system, and a cross-over toward the sphere-cube geometry is expected. We also show in Fig. \ref{varz_CP} the standard deviation of the CP potential relative to its mean, $\gamma$, as a function of $nz^3$. Red dots are the numerical results, and the dashed curve is the theoretical prediction (\ref{gamma_theo}).
The agreement between theory and numerics is very good, up to small finite-size effects at large $z$.

\subsection{Probability distribution function}
\label{ps_numerics}

\begin{figure}
\includegraphics[width=0.8\linewidth]{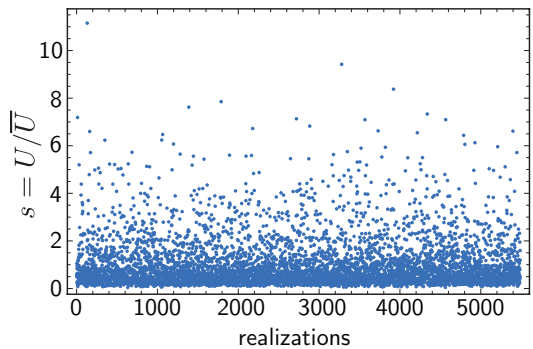}
\caption{
\label{scattered_U}
(color online) Scatter plot of the CP potential normalized to its mean value, $s=U/\overline{U}$, computed for 5500 disorder realizations, for $nz^3=0.5$.
}
\end{figure}

\begin{figure}
\includegraphics[width=1.0\linewidth]{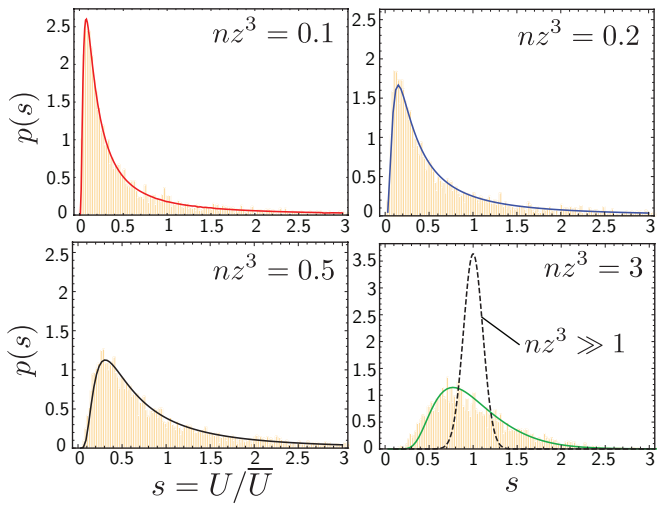}
\caption{
\label{Ps_fig}
(color online) Probability distribution function $p(s=U/\overline{U})$, for four increasing  values of $nz^3$. Histograms (vertical lines) are the results of exact numerical simulations, and solid curves are the theoretical prediction, Eq. (\ref{ps_general}). The lower-right panel also displays as a dashed curve the Gaussian distribution expected in the limit of very large distances, Eq. (\ref{gaussian_ps}).
}
\end{figure}

As was pointed out in \cite{Cherroret15}, the fluctuations of $U(z)$ become significant at distances $z\lesssim n^{-1/3}$, when $\gamma$ becomes larger than unity, see Eq. (\ref{gamma_theo}) and Fig. \ref{varz_CP}. This suggests that at small distances, the mean $\overline{U}(z)$ is no longer representative of $U(z)$. In order to confirm this picture, we compute the full probability distribution function $p(s)$ of the CP potential normalized to its mean, $s=U/\overline{U}$, by constructing histograms of the numerical data. In Fig. \ref{scattered_U}, we show as an example a scatter plot of the data obtained for 5500 disorder realizations, for $nz^3=0.5$. The associated histogram $p(s)$ is displayed in Fig. \ref{Ps_fig}, together with the histograms corresponding to three other values of the parameter $nz^3$.
A quick look at the distributions in Fig. \ref{Ps_fig} forthwith confirms the property already outlined by the analysis of $\gamma$: as the sphere gets closer to the disordered medium, the distribution function becomes more and more peaked around a value $s\ll 1$, corresponding to a CP potential much smaller than its mean. In other words, $U(z)$ is no longer a self-averaging quantity. Only when $nz^3>1$ does the maximum of the distribution approaches $s=1$. Such a phenomenon was previously observed in the context of the interaction between plates with one-dimensional disorder \cite{Dean09, Dean10}. We see here that it is a quite general property, not restricted to one-dimensional systems.

\subsection{Sensitivity to microscopic parameters}
\label{universality}

To conclude our numerical study, we address the question of the sensitivity of $p(s)$ with respect to changes in the microscopic properties of the scatterers. For this purpose, we perform additional numerical simulations involving scatterers with a different radius $a=1\,$nm and made of a different material with frequency-independent permittivity $\epsilon=10$. We also set the radius of the probe sphere to $a=1\,$nm, but keep the same value of the permittivity of silicon as in the previous section.
\begin{figure}[h]
\begin{center}
\includegraphics[scale=.51]{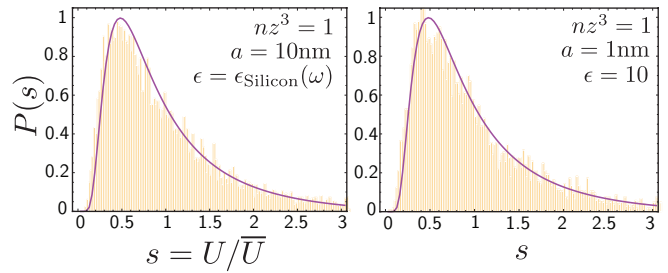}
\caption{\label{universality_fig}
(color online) Numerical probability distribution $p(s)$ for two different sets of parameters of the scattering spheres, for $nz^3=1$ (histograms, vertical lines). No visible difference is seen between the two histograms. This is confirmed by the theoretical prediction (solid curve), Eq. (\ref{ps_general}), which depends on the single parameter $nz^3$.
}
\end{center}
\end{figure}
The distribution $p(s)$ obtained for these new parameters is shown in the right panel of Fig. \ref{universality_fig}, for $nz^3=1$. For comparison, we also display the distribution $p(s)$ computed with the parameters of the previous section. No visible difference is seen between the two histograms, which indicates that in the dilute limit $p(s)$ is in fact a function of the parameter $nz^3$ only. In particular, the parameter $na^3$ is irrelevant. This could have been anticipated since in the limit $na^3\ll 1$ of independent scatterers, $na^3$ enters both $U$ and $\overline{U}$ within the same prefactor, which thus cancels out when considering the ratio $s=U/\overline{U}$ (see Sec. \ref{theory_sec} for a general proof). This property is in particular fulfilled by the second moment of the distribution, $\gamma^2=\overline{\delta U^2}/\overline{U}^2$, see Eq. (\ref{gamma_theo}).

\section{Simple model}
\label{theory_sec}

We now develop a simplified statistical description of the CP interaction between a dielectric sphere and a disordered bulk medium, based on a pairwise summation (PWS) approximation \cite{Bitbol13}.
This approximation describes the total CP interaction $U(z)$ as a sum of the pair interaction $\mathcal{E}$ between the probe sphere and each of the $N$ scatterers. 
It has to be distinguished from the perturbative expansion, as it can in principle be used for non perturbative pair interactions.
In the problem studied in this paper however, the validity of the two approximations is a consequence of the same assumption of a dilute disorder ($na^3\ll 1$).

\subsection{Pairwise summation}
\label{PWS}

\begin{figure}
\includegraphics[width=0.92\linewidth]{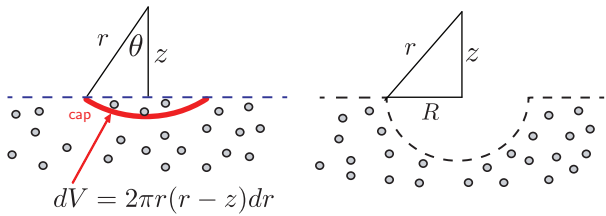}
\caption{
\label{parametrization}
(color online) Left: parametrization of the geometry for the statistical approach of Sec. \ref{PWS}. Right: typical (rare) disorder realization contributing to the Lifshitz tail: the atom is above a large region free of scatterers, of size $R\gg z$.
}
\end{figure}

As in the numerical simulations (Sec. \ref{numerics_sec}), we consider a situation where the distance $z$ much exceeds the sphere resonance wavelength $\lambda_0$ (In the opposite limit $z\ll\lambda_0$, the reasoning follows exactly the same lines). Consequently, the interaction potential between the probe sphere and a scatterer located at distance $r$ takes the simple form $\mathcal{E}=-\Gamma_7/r^7$, where $\Gamma_7$ is a constant characteristic of the microscopic properties of the two interacting objects. Let us consider a small spherical cap of volume $dV$ containing $dN$ scatterers, as illustrated in the left panel of Fig. \ref{parametrization}. The elementary CP interaction between them and the probe sphere is $dU=\mathcal{E} dN=-\Gamma_7dN/r^7$. As in the simulations, we assume the positions of the scatterers to follow a uniform distribution, so the random variable $dN$ is Poisson distributed. The $m^\text{th}$ cumulant of $dN$, $K_m(dN)$, thus fulfills:
\begin{equation}
\label{KmdN}
K_m(dN)=K_1(dN)=ndV,
\end{equation}
where $n=N/L^3$ is the average density of scatterers. Within the PWS approximation, the cumulant of the elementary CP potential $dU$ involving $dN$ scatterers reads:
\begin{eqnarray}
K_m(dU)&=&\left(-\frac{\Gamma_7}{r^7}\right)^m K_m(dN)\nonumber\\
&=&\left(-\frac{\Gamma_7}{r^7}\right)^m ndV\equiv d K_m(U),
\end{eqnarray}
where we have used the property $K_m(aX)=a^mK_m(X)$ in the first equality, and Eq. (\ref{KmdN}) in the second. The cumulant of the total CP potential between the probe sphere and the $N$ scatterers is finally obtained  by using the parametrization $dV=2\pi r(r-z)dr$, see Fig. \ref{parametrization}, and integrating over $r$ from $r=z$ to $\infty$.
This yields
\begin{equation}
\label{cumulants}
K_m(U)=\frac{(-1)^m 2\pi n\Gamma_7^m}{(7m-3)(7m-2)z^{7m-3}},
\end{equation}
from which various statistical properties can be deduced, as we now discuss.

\subsection{Mean and standard deviation}
\label{mean_var_PWS}

As a preliminary application of our statistical approach, we propose to re-derive the predictions (\ref{Ubar_theo}) and (\ref{gamma_theo}), previously obtained from an exact treatment of the radiation-matter interaction. For simplicity we assume the scatterers to be spheres of static polarizability $\alpha_0$, identical to the probe sphere. The coefficient $\Gamma_7$ then describes the large-distance interaction between two identical spheres. It can be readily evaluated, for instance from the Casimir-Polder law \cite{Casimir48} for the energy between two atoms \cite{Power93}:
\begin{equation}
\label{gammasept}
\Gamma_7=\frac{23\hbar c\alpha_0^2}{(4\pi)^3}.
\end{equation}

The mean $\overline{U}(z)$ is by definition the first-order cumulant: $\overline{U}(z)=K_1(z)=-2\pi n\Gamma_7/(20z^4)$, where we have used Eq. (\ref{cumulants}) for $m=1$. Combining this expression with Eq. (\ref{gammasept}), we recover Eq. (\ref{Ubar_theo}). The variance $\overline{\delta U^2}(z)$ is, on the other hand, given by the second-order cumulant, $K_2(z)=2\pi n\Gamma_7^2/(132 z^{11})$. Taking the ratio with $K_1(z)$, we find
\begin{equation}
\label{gamma_theo_PWS}
\gamma=\frac{\sqrt{\overline{\delta U^2}(z)}}{|\overline{U}(z)|}\simeq\sqrt{\frac{50}{33\pi}}\frac{1}{\sqrt{nz^3}},
\end{equation}
which is nothing but Eq. (\ref{gamma_theo}), with an analytic expression of the coefficient $\sqrt{50/(33\pi)}\simeq 0.7$.

\subsection{Probability distribution function}

Let us now derive the probability distribution function $p(s=U/\overline{U})$ that has been studied numerically in Sec. \ref{ps_numerics}. $p(s)$ is given by the inverse Laplace transform
\begin{equation}
\label{Lmoinsun}
p(s)=\dfrac{1}{2\pi i}\int_{\delta-i\infty}^{\delta+i\infty}
e^{st}e^{\varphi(t)}dt,
\end{equation}
where $\delta$ is greater than the real part of all singularities of $e^{\varphi(t)}$. $\varphi(t)$ is the cumulant generating function of $s$, and can be expressed as a power series of the cumulants (\ref{cumulants}):
\begin{equation}
\varphi(t)=\sum_{m=0}^\infty
\dfrac{(-t)^m}{m!}\frac{K_m(U)}{\overline{U}^m}.
\end{equation}
Making use of Eq. (\ref{cumulants}) and of the relation $\overline{U}(z)=-2\pi n\Gamma_7/(20z^4)$ obtained above, we find:
\begin{equation}
\label{phit}
\varphi(t)=\dfrac{2\pi n z^3}{6}
\left[-1+e^{-\tau}-2\tau^{3/7}\gamma_{4/7}(\tau)+3\tau^{2/7}\gamma_{5/7}(\tau)
\right],
\end{equation}
where we have introduced $\tau=20t/(2\pi n z^3)$ and where $\gamma_{q}(\tau)=\int_0^\tau x^{q-1}e^{-x}dx$ is the lower incomplete gamma function. 
Since $e^{\varphi(t)}$ has no singularities in the complex plane, we can set $\delta=0$ in Eq. (\ref{Lmoinsun}). Furthermore, we have the property $\varphi^*(t)=\varphi(t^*)$, such that after the substitution $t=ix$, Eq. (\ref{Lmoinsun}) simplifies to 
\begin{equation}
\label{ps}
p(s)=\dfrac{1}{\pi }\int_{0}^{\infty}
\text{Re}\left[e^{i x s}e^{\varphi(i x)}\right]dx.
\end{equation}
Inserting Eq. (\ref{phit}) into this relation, we finally obtain
\begin{widetext}
\begin{equation}
\label{ps_general}
p(s)=\dfrac{nz^3}{10}\text{Re}\int_{0}^{\infty}
\exp\left\{\dfrac{2\pi n z^3}{6}\left[
\dfrac{3}{10}s i \tau-1+e^{-i\tau}-2(i\tau)^{3/7}\gamma_{4/7}(i\tau)+3(i\tau)^{2/7}\gamma_{5/7}(i\tau)
\right]
\right\}d\tau.
\end{equation}
\end{widetext}
Distributions $p(s)$ obtained from numerical evaluation of Eq. (\ref{ps_general}) are displayed in Fig. \ref{Ps_fig} as solid curves on top of the numerical results of Sec. \ref{numerics_sec}. The agreement is excellent for all the values of $nz^3$. Furthermore, we notice that Eq. (\ref{ps_general}) confirms the conclusion drawn from the simulations in Sec. \ref{universality}: $p(s)$ depends on the parameter $nz^3$ only, being completely independent of the microscopic details of the probe sphere and of the scatterers.

\subsection{Asymptotics}

Although the distribution $p(s)$, Eq. (\ref{ps_general}), has no evident analytic expression, several simple asymptotic limits can be readily examined.\newline

\noindent \emph{Large-$s$ limit}
We describe the limit of large $s$ by expanding the term inside the square brackets in Eq. (\ref{ps_general}) up to second order in $\tau\ll1$. Since essentially the values of $\tau$ such that $\tau s nz^3 \lesssim1$ contribute to the integral, this expansion is a good approximation provided $1/(snz^3)\ll1$. It results in a Gaussian integral which is straightforwardly performed to give:
\begin{equation}
\label{gaussian1}
p(s)\simeq\dfrac{\sqrt{33nz^3}}{10}
\exp\left[
-\dfrac{33\pi}{100}nz^3(s-1)^2
\right].
\end{equation}
Making use of Eq. (\ref{gamma_theo_PWS}), we rewrite Eq. (\ref{gaussian1}) as 
\begin{equation}
\label{gaussian_ps}
\dfrac{1}{\sqrt{2\pi\overline{\delta U^2}}}\exp\left[-\dfrac{(U-\overline{U})^2}{2\overline{\delta U^2}}\right].
\end{equation}
At large $s$, $p(s)$ is thus simply a (normalized) Gaussian distribution of mean $\overline{U}$ and variance $\overline{\delta U^2}$. 
The expansion used to derive Eq. (\ref{gaussian_ps}) being valid as long as $s\gg (nz^3)^{-1}$, the Gaussian shape is a very good approximation of the whole distribution when $nz^3\gg1$, which is a direct consequence of the weakness of fluctuations at large distances. 
%Indeed, when $nz^3\gg1$ all cumulants of order greater than three become negligible. 
We expect this Gaussian distribution to be universal at large distances, as a consequence of the central-limit theorem (many scatterers contribute to $U(z)$ when $nz^3\gg1$), regardless the nature of the disordered medium. This conclusion is supported by similar predictions previously made in the context of the thermal Casimir effect in one-dimensional disordered media \cite{Dean09, Dean10}, as well as in recent studies of the CP interaction involving quasi two-dimensional disordered metals \cite{Allocca15}. For comparison, we show Eq. (\ref{gaussian_ps}) in the lower-right panel of Fig. \ref{Ps_fig} as a dashed curve, for $nz^3=40$. Note that in the chosen geometry this limit is difficult to reach in the numerical simulations, because it requires the generation of a significant number of scattering spheres in order to satisfy the condition $nz^3\gg 1$, while maintaining $z\lesssim L/2$ to avoid finite-size effects. When $nz^3\ll 1$, Eq. (\ref{gaussian_ps}) still holds but only in the very far tail of the distribution, $s\gg(nz^3)^{-1}\gg 1$. The physical reason for which we recover a Gaussian tail even for small values of $nz^3$ is the following. Large values of $s$  correspond to particular disorder realizations $\mathcal D$ for which the Casimir potential is very large, i.e. for which the density of scatterers $n_{\mathcal D}$ underneath the probe atom is very high. Thus, for these specific disorder realizations the condition $n_{\mathcal D} z^3\gg 1$ is effectively fulfilled.\newline

\noindent\emph{Moderate values of $s$}
\begin{figure}
\includegraphics[scale=0.7]{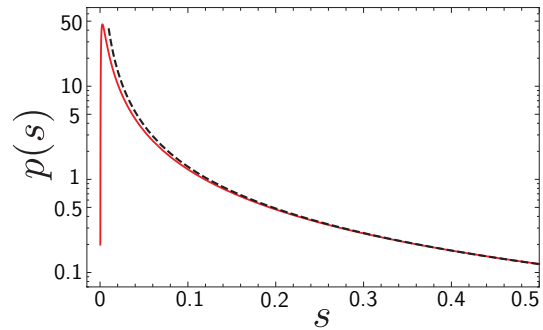}
\caption{
\label{ps_moderates}
(color online) Distribution $p(s)$ for $nz^3=0.005$ (solid red curve), together with the approximate form (\ref{ps_moderate}) at moderate $s$ (dashed curve).
}
\end{figure}
To describe the limit $s\ll (nz^3)^{-1}$, we expand the term inside the square brackets in Eq. (\ref{ps_general}) for large $\tau$:
\begin{eqnarray}
\label{ps_smallnzc}
&&p(s)\simeq\dfrac{nz^3}{10}\text{Re}\int_{0}^{\infty}
d\tau
\exp\left\{
\dfrac{2\pi nz^3}{6}
\left[
\dfrac{3}{10}i s\tau-2\Gamma(4/7)
\right.\right.\nonumber\\
&&
\left.\left.\times \left(i\tau\right)^{3/7}\!\!\!
+3\Gamma(5/7)\left(i\tau\right)^{2/7}\right]\right\}.
\end{eqnarray}
We then perform a Wick rotation $\tau=ix$, and express both exponential terms $\exp(\alpha x^{3/7})$ and $\exp(\beta x^{2/7})$ as power series. This gives
\begin{eqnarray*}
\label{ps_series}
&&p(s)=-\dfrac{nz^3}{10}\text{Im}\int_{0}^{\infty}
dx
\exp\left(-\dfrac{\pi nz^3}{10}s x\right)
\sum_{n,m=0}^\infty\dfrac{1}{n! m!}\\
&&
\times\left[-\dfrac{2\pi n z^3}{3}\Gamma(4/7)\left(-x\right)^{3/7}\right]^n
\left[\pi n z^3\Gamma(5/7)\left(-x\right)^{2/7}\right]^m.
\end{eqnarray*}
If we additionally assume $s\gg n z^3$, the terms $(m,n)=(0,1)$ and $(m,n)=(1,0)$ give the leading contribution to $p(s)$ (the term $m=n=0$ is purely real and does not contribute). Keeping only these two terms and computing the remaining integral, we find
\begin{equation}
\label{ps_moderate}
p(s)\simeq\dfrac{2}{7}10^{3/7}
\dfrac{(\pi n z^3)^{4/7}}{s^{10/7}}\left[1-\left(\dfrac{\pi n z^3 s}{10}\right)^{1/7}\right].
\end{equation}
Equation (\ref{ps_moderate}) holds for $nz^3\ll s\ll (nz^3)^{-1}$. It is shown in Fig. \ref{ps_moderates} as a dashed curve for $nz^3=0.005$, together with the exact distribution calculated from Eq. (\ref{ps_general}) (solid red curve).\newline

\noindent \emph{Small-$s$ limit}
We finally consider the low-potential tail $s\ll nz^3$ of $p(s)$. In order to find an asymptotic expansion in that limit, we come back to Eq. (\ref{ps_smallnzc}) and apply the change of variables $\tau=(y/s)^7$. This yields
\begin{eqnarray}
p(s)&\simeq&\dfrac{7nz^3}{10s^7}\text{Re}\int_{0}^{\infty}
y^6dy
\exp\left\{
\dfrac{2\pi nz^3}{6s^6}
\left[
\dfrac{3}{10}i y^7
\right.\right.\nonumber\\
&&\left.\left.
-2\Gamma(4/7)i^{3/7}s^3y^3
+3\Gamma(5/7)i^{2/7}s^4y^2\right]\right\}.
\end{eqnarray}
In the limit of very small $s$, we are thus led to evaluate 
\begin{equation}
\label{I_def}
I=\text{Re}\int_0^\infty dy y^6\exp\left[\Lambda f(y)\right],
\end{equation}
where $\Lambda=\pi n z^3/(3 s^6)\gg1$ and $f(y)=(3/10)i y^7-2\Gamma(4/7)i^{3/7}s^3y^3+3\Gamma(5/7)i^{2/7}s^4y^2$. Equation (\ref{I_def}) naturally calls for the method of steepest descent. There are five saddle points, solutions of $f'(y)=0$. 
%These saddle points can be calculated by solving the equation $f'(y)=0$ using perturbation theory in the limit $s\rightarrow 0$. 
Only one of them, denoted by $y_\text{SP}$, turns out to give a nonzero contribution to $p(s)$:
\begin{equation}
y_\text{SP}=\left[\dfrac{60\Gamma(4/7)s^3}{21}\right]^{1/4}e^{-i\pi/14}.
\end{equation}
Using Cauchy's theorem, we then deform the path of integration to a path coinciding with the path of steepest descent in the vicinity of $y_\text{SP}$. This is achieved by expanding $f(y)$ up to second order around $y=y_\text{SP}$ and performing the change of variables $x=(y-y_\text{SP})\exp(-3i\pi/14)$, leaving us with a Gaussian integral whose evaluation leads to:
%\begin{equation}
%I\simeq\int_{-\infty}^{\infty}dx |y_\text{SP}|^6\exp\left[\Lambda f(y_\text{SP})\right]\exp\left[-\Lambda \dfrac{|f''(y_\text{SP})|}{2}x^2\right].
%\end{equation}
%Performing the Gaussian integral, we finally obtain:
\begin{eqnarray}
\label{Lifshitz_tail}
p(s)\simeq\alpha\dfrac{\sqrt{nz^3}}{s^{11/8}}
&&\exp\left[-\beta\dfrac{nz^3}{s^{3/4}}\right],
\end{eqnarray}
with prefactors $\alpha=(20/7)^{3/8}\Gamma(4/7)^{7/8}$ and $\beta=16\sqrt{2}\pi(5/7)^{3/4}\Gamma(4/7)^{7/4}/21$. The asymptotic form (\ref{Lifshitz_tail}) is completely analogous to the so-called Lifshitz tail that describes the band edge of the density of states of disordered conductors in solid-state physics \cite{Lifshitz1964}. Physically, it can be understood from the following qualitative argument. Low values of $s$ are achieved for rare disorder realizations where the probe sphere stands above a large region free of scatterers, as illustrated in the right scheme of Fig. \ref{parametrization}. Since the distribution of scatterers is Poissonian (Sec. \ref{theory_sec}), the distance between the events of this Poisson process follows an exponential distribution. Consequently, the probability  to find a large region of size $R$ free of scatterers is $\propto\exp(-cn R^3)$, where $c$ is a numerical constant. In  such configuration, the Casimir potential felt by the atom can be estimated as (see Fig. \ref{parametrization}):
\begin{equation}
U=-2\pi n\Gamma_7\int_{\sqrt{z^2+R^2}}^\infty
\dfrac{r(r-z)dr}{r^7}
\propto 
\dfrac{-n\Gamma_7}{R^4},
\end{equation}
to leading order in $z/R\ll 1$. On the other hand, we have seen in Sec. \ref{mean_var_PWS} that the average Casimir potential is $\overline{U}\propto -n\Gamma_7/z^4$. Therefore, for the rare disorder realization displayed in Fig. \ref{parametrization}, we have $U\propto (z/R)^4 \overline{U}$, such that
\begin{eqnarray}
p(s)&\sim&\exp(-cn R^3)=\exp\left[-cn \left(z\dfrac{\overline{U}^{1/4}}{U^{1/4}}\right)^3\right]\nonumber\\
&=&\exp\left(-\dfrac{cnz^3}{s^{3/4}}\right),
\end{eqnarray}
which is nothing but the asymptotic form (\ref{Lifshitz_tail}).\newline

\section{Conclusion}
\label{conclusion_sec}

In this paper, we have developed a statistical description of Casimir-Polder potentials from both a numerical and an analytical perspective. This approach is well suited for describing the Casimir interaction between a simple dielectric object and a strongly heterogenous medium made of a large number of independent constituting elements. It can be readily extended to other geometries and to heterogeneous media characterized by a more complex statistics involving, for instance, a non-uniform or polydisperse distribution of scatterers.

As a first extension of our work, it would be interesting to investigate deviations to the dilute limit $n a^3\ll 1$ where the scatterers can no longer be systematically considered independent. We expect these deviations to primarily affect the far tails of the distribution $p(s)$. A second open question concerns the change in the statistics of the Casimir-Polder potential when the host (homogeneous) medium has a dielectric constant differing from unity. This problem is more difficult to treat since it now involves a surface, which implies multiple reflections inside and outside the medium. The presence of a host medium may therefore strongly affect the distribution $p(s)$ [in the obvious limit where the dielectric constant of the host medium goes to infinity, one recovers a perfectly reflecting interface and $p(s)$ should tend to the Dirac function $\delta(s-1)$].

In practice, the distribution $p(s)$ could be experimentally accessed either by moving the sphere over a static disordered medium to record different disorder distributions, or by taking advantage of a Brownian motion of the scatterers if the measurement process is fast enough. Indeed, in that case different disorder realizations can be obtained by detecting the Casimir force and then letting the scatterers move before carrying out the next measurement. If the measurement process is slow, the effect of the motion of the scatterers is to average the Casimir potential, giving him a value well approximated by Eq. (\ref{Ubar_theo}) since Doppler shifts have a negligible effect at thermal velocities \cite{Ferrell80, Barton10, Dedkov11}.

\appendix

\end{document}